\documentclass[prd,twocolumn,amsmath,amssymb]{revtex4-1}
\usepackage{amsmath}
\usepackage[caption=false]{subfig}
\usepackage{bbm}
\usepackage[utf8x]{inputenc}

\usepackage{mathtools}
\usepackage{bbold}

\usepackage{appendix}
\usepackage{xcolor}
\usepackage{float}
\usepackage{braket}
\usepackage{slashed}
\usepackage{graphicx}
\definecolor{darkred}{rgb}{0.4,0.0,0.0}
\definecolor{darkgreen}{rgb}{0.0,0.3,0.0}
\definecolor{darkblue}{rgb}{0.0,0.0,0.7}
\usepackage[bookmarks,linktocpage,colorlinks,
linkcolor = darkred,
urlcolor  = darkgreen,
citecolor = darkblue]{hyperref}
\def\beq{\begin{equation}}
  \def\enq{\end{equation}}
\usepackage{lineno}
\usepackage{comment}
\definecolor{winered}{rgb}{0.8,0,0}
\definecolor{darkb}{rgb}{0,0,0.8}

\usepackage[english]{babel}

\def\rar{\rightarrow}

\def\mc{\mathcal{C}}
\def\mcm{\mathcal{C}^{-1}}
\begin{document}

\title{Theoretical methods to design and test quantum simulators for  the compact Abelian Higgs model} 
\author{Yannick Meurice$^{1}$}
\affiliation{$^1$ Department of Physics and Astronomy, The University of Iowa, Iowa City, IA 52242 USA }
\def\lt{\lambda ^t}
\def\note{note}
\def\beq{\begin{equation}}
  \def\enq{\end{equation}}

\date{\today}

\begin{abstract}
The lattice compact Abelian Higgs model is a non-perturbative regularized formulation of low-energy scalar quantum electrodynamics. In 1+1 dimensions, this model can be quantum simulated using a ladder-shaped optical lattice with Rydberg-dressed atoms  \cite{prl121}. In this setup, one spatial dimension is used to carry the angular momentum of the quantum rotors.
One can use truncations corresponding to spin-2 and spin-1 to build local Hilbert spaces associated with the links of the lattice. 
We argue that ladder-shaped configurable arrays of Rydberg atoms can be used for the same purpose. We make concrete proposals involving two and three Rydberg atoms to build one local spin-1 space (a qutrit). 
We show that the building blocks of the Hamiltonian calculations are models with one and two spins. 
We compare target and simulators using perturbative and numerical methods. The two-atom setup provides an easily controllable simulator of the one-spin model while the three-atom setup involves solving nonlinear equations. 
We discuss approximate methods to couple two spin-1 spaces. 
The article provides analytical and numerical tools necessary to design and build the proposed simulators with 
current technology. 
    \end{abstract}

\maketitle

\section{Introduction}

\def\cara{CARA}
There has been recent interest in using quantum simulations and quantum computations to address problems with real-time and finite density in high-energy physics \cite{banuls2020,Klco:2021lap,Aidelsburger:2021mia,Wiese:2021djl,Gustafson:2021imb,Lamm:2019bik,Gustafson:2020yfe,Brower:2020huh,Zohar:2015hwa,Wiese:2013uua}.
One initial step is the simulation of Abelian gauge theories \cite{Zohar:2011cw,Tagliacozzo:2012vg}. 
The Schwinger model introduces fermions and can be studied with methods developed in many-body physics \cite{Banuls:2013jaa,Buyens:2015tea,Banuls:2016lkq,Funcke:2019zna}. Quantum simulations \cite{Martinez:2016yna,Kasper:2016mzj} and quantum computations \cite{Klco:2018kyo,Kharzeev:2020kgc} have been performed for this model. 

A bosonic variant is the compact Abelian Higgs model. The compactness allows discrete character expansions formulations \cite{prd92,prd98,rmp}  which are gauge-invariant and solve \cite{prd102} the questions of gauge redundancy and the implementation of Gauss's law \cite{Unmuth-Yockey:2018xak,Bender:2020ztu}. The truncations do not break symmetries \cite{prd100,prd102,rmp} but can affect the nature of phase transitions \cite{Zhang:2021dnz,Hostetler:2021uml}. 
The non-compact Brout-Englert-Higgs mode is assumed to be decoupled hereafter. For a recent discussion of its effects 
in the context of quantum simulations see Ref. 
\cite{titas}. 
An Optical lattice implementation with spin-2 
has been proposed \cite{prl121} in 1+1 dimensions. 
It is based on  a ladder-shaped optical lattice with Rydberg-dressed atoms. In this setup, one spatial dimension carries the angular momentum of quantum rotors.
Truncations corresponding to local Hilbert spaces with various spin truncations associated with the links of the lattice have been discussed \cite{prd92,prd98,prl121,Zhang:2021dnz}.

In the following, we argue that ladder-shaped configurable arrays of Rydberg atoms  \cite{51qubits,endres2016,keesling2019,cong2021hardwareefficient,semeghini2021probing}, abbreviated CARA, can be used for the same purpose. This platform has been used for other lattice gauge theory models \cite{PhysRevX.10.021041,Celi:2019lqy,PhysRevResearch.2.013288}. See Ref. \cite{Wu:2020axb} for a review of the use of Rydberg atoms. 
We make concrete proposals involving two and three Rydberg atoms to build one local spin-1 space (a qutrit). 
We show that the building blocks of the Hamiltonian calculations are simple models with one and two spins. 
We compare target and simulators for these simple models using perturbative and numerical methods. The two-atom setup provides an easily controllable simulator of the one-spin model while the three-atom setup involves nonlinear matching which could be tested with 
current technology. We argue that near-term technology could be used to quantum simulate models with two or more spins.
More generally, programming with CARA amounts to a geometrical assembling allowing a broad range of applications. 
The idea of providing qutrits is very timely \cite{Ciavarella:2021nmj,Gustafson:2021qbt}.

The article is organized as follows. In Sec. \ref{sec:model}, we review the Lagrangian and Hamiltonian formulation of the 
compact Abelian Higgs model with emphasis on the meaning of the signs of the couplings. The general idea of ladder-shaped 
CARA is introduced in Sec. \ref{sec:simul}. The two and three atoms CARA for a single spin-1 are discussed  in Sec. \ref{sec:onespin}. The coupling 
of two spins with an operator that is the product of their respective angular momentum $L^z$'s is discussed in Sec. \ref{sec:twospins}.
We argue that the single spin-1 simulators can be tested with
current technology and  that near-term technology could be used to quantum simulate models with two or more spins. 
Implementations with universal quantum computers are discussed in Sec. \ref{sec:ibm} and the conclusions are provided in Sec. \ref{sec:conclusions}.

\section{The lattice model}
\label{sec:model}
In this section we review the Lagrangian and Hamiltonian formulations of the compact Abelian Higgs model with emphasis on the meaning of the signs of the couplings. As we will see the sign question is important from the point of view of quantum simulations.
\subsection{Lagrangian formulation}
We first review the Lagrangian path integral formulation of the Abelian Higgs model at Euclidean time
using most of the notations of Ref. \cite{prd92} which should be consulted for more details.  
The partition function has the form
\begin{equation}
Z=\int D\phi^\dagger D\phi DU e^{-S} .\end{equation}
The action is the sum of three terms
\begin{equation}\label{S_u1h}
S = S_g + S_h + S_\lambda,
\end{equation}
where the gauge part is
\begin{equation}
S_g = -\beta_{pl}\sum_x{\rm Re}\left[U_{pl.,x}\right],
\end{equation}
the hopping part
\begin{eqnarray}
S_h &=& -{\kappa_\tau}\sum_x
\left[\phi_x^\dagger U_{x,\hat\tau}\phi_{x+\hat\tau}+
\phi_{x+\hat\tau}^\dagger U^\dagger_{x,\hat\tau}\phi_x
\right] \nonumber\\ 
&-&{\kappa_s}\sum_x
\left[\phi_x^\dagger U_{x,\hat{s}}\phi_{x+\hat{s}}+
\phi_{x+\hat{s}}^\dagger U^\dagger_{x,\hat{s}}\phi_x
\right],
\end{eqnarray}
and the self-interaction
\begin{equation}
S_\lambda = \lambda\sum_x\left(\phi_x^\dagger\phi_x-1\right)^2+
\sum_x\phi_x^\dagger\phi_x .
\end{equation}
By writing 
\beq
\phi_x=|\phi_x|\exp(i\varphi_x),
\enq
we can separate the compact and non-compact variables in $S_h$:
\begin{eqnarray}
S_{h}=&-&  2\kappa_\tau |\phi_x||\phi_{x+\hat\tau}| \sum\limits_{x} \cos(\varphi_{x+\hat\tau} - \varphi_{x}+A_{x,\hat\tau})\cr&-&
2\kappa_s |\phi_x||\phi_{x+\hat{s}}| \sum\limits_{x} \cos(\varphi_{x+\hat{s}} - \varphi_{x}+A_{x,\hat{s}}) .
\end{eqnarray}
It is assumed that $\kappa_s$ and $\kappa_\tau$ are positive as in ferromagnetic interactions. This means that if we neglect the gauge fields and the self-interactions, large values of $|\phi|$ favor the alignment of the matter fields (configurations where all the angles $\varphi_x$ are equal), as expected in the continuum limit of the free $O(2)$ scalar model. 

In the following, we take the limit where $\lambda$ become large and positive. The Brout-Englert-Higgs mode $|\phi |$ is then frozen to 1. The Nambu-Goldstone mode $\varphi$ is compact. 
We call this model the compact Abelian Higgs model . 
By shifting the integration variable $\varphi$ by $\pi$ on every other site say in the spatial direction, we can flip the sign of $\kappa_s$ without affecting the partition function. 
A similar reasoning can be applied for $\kappa_{\tau}$ and the time direction. However, if observable are inserted in the partition function, these changes of variable need to be performed for the observables too. For instance, the magnetization becomes a staggered magnetization. Similar considerations apply to the plaquette term and the sign of $\beta_{pl.}$. 
This is discussed at length in \cite{li2004,prd80}.

\subsection{Hamiltonian and Hilbert space}
Following Ref. \cite{prd92,prl121,prd98} the continuous-time limit for the compact Abelian Higgs model was taken in the field quantum number representation in the limit where the Higgs quartic self-coupling goes to infinity.  
To take the time continuum limit, one takes $\kappa_{\tau}, \beta_{pl} \rightarrow \infty$ while simultaneously taking $\kappa_{s}$, and the temporal lattice spacing, $a$, to zero such that the combinations
\begin{equation}
\label{eq:abhamil}
	U \equiv \frac{1}{\beta_{pl} a} = \frac{g^{2}}{a}, \quad
    Y \equiv \frac{1}{2 \kappa_{\tau} a}, \quad
    X \equiv \frac{2 \kappa_{s}}{a}
\end{equation}
are finite.   These equations  make clear that the signs of $U$, $X$ and $Y$ are the same as $\beta_{pl.}$, $\kappa_{\tau}$ and $\kappa_s$ respectively. 
The Hamiltonian for $N_s$ links reads
\begin{align}
	\label{eq:ham}
	H &= \frac{U}{2}\sum_{i=1}^{N_{s}} \left(L^z_{i}\right)^2 \nonumber \\
	&+ \frac{Y}{2} {\sum_i}  ' (L^z_{i+1} - L^z_{i})^2-
	X\sum_{i=1}^{N_{s}} U^x_{i} \ ,
\end{align}
where the sum, $\sum_i '$, means that  for  open boundary conditions (OBC) we need to include $(L^z_1)^2 + (L^z_{N_s})^2$. 
We used the operator 
\beq
U^x\equiv \frac{1}{2}(U^+ + U^-), 
\enq
with
\beq
U^\pm\ket{m}=\ket{m\pm1}.
\enq
The quantum number $m$ corresponds to the Fourier modes in the character expansion of the Lagrangian formulation \cite{prd92}. In practice, we need to apply  truncations.  
For a spin-$m_{max}$ truncation we have 
\beq
U^\pm\ket{\pm m_{max}}=0.
\enq
In the following we mostly focus on the spin-1 truncation where $m = \pm 1, 0$. In this special case $U^x=L^x/\sqrt{2}$.

The truncations are compatible with the identities following from local or global symmetries of several models with continuous Abelian symmetries \cite{prd100,prd102,rmp}, however they can affect the type of phase transition present in the 
the model. 

The physical interpretation of the three terms of the Hamiltonian is the same as in conventional electrodynamics except for the fact that all the values involved are discrete.
The $U$-term represents the electric field energy.
The $Y$-term is associated with matter charges. They can be interpreted as charges determined by Gauss's 
law, in other words the difference between the two plaquettes (electric field) on each side of a site in the Lagrangian form. 
Finally the $X$-term is related to matter currents inducing temporal changes in the electric field, again in the Lagrangian form. This corresponds to the other inhomogeneous Maxwell equation involving the currents. In higher dimensions, the 
discrete curl of a magnetic field appears as in the continuum \cite{prd102}.

\subsection{Charge conjugation}

As in standard quantum electrodynamics this Hamiltonian has a charge conjugation symmetry. 
This will play an important role in the construction of simulators because this property will translate into a global reflection 
symmetry in the geometrical setup of the atoms. For this reason we remind the basic equations associated with this symmetry. The charge conjugation $\mc$ is a unitary transformation which reverses the sign of $m$:
\beq 
\mc\ket{m}=\ket{-m}
\enq
It is clear that $\mc^2=1$ and that 
\begin{align}
&\mc L^z \mcm =-L^z,\\ & \mc U^\pm \mcm = U^\mp,\\
&\mc U^x \mcm = U^x
\end{align}
This implies that the Hamiltonian is invariant under charge conjugation:
\beq
\mc H\mcm = H.
\enq

\subsection{The building blocks of the Hamiltonian formalism}

In order to conduct actual experiments to quantum simulate the compact Abelian Higgs model, 
we need to identify its building blocks. The first one is the local spin-1 or higher spin Hilbert space where we need to 
implement the operators $L^z$  and $U^x$. This will be discussed in Sec. \ref{sec:onespin}. The second is the coupling 
of two spins with an operator that is the product of their respective $L^z$'s. This will be discussed in Sec. \ref{sec:twospins}.

\section{Rydberg atom simulators}
\label{sec:simul}

\subsection{A ladder-shaped optical lattice simulator}
\label{subsec:ladder}

In Ref. \cite{prl121}, a quantum simulator for the spin-2 truncation of the Hamiltonian in Eq. (\ref{eq:ham}) was proposed.
The general idea is to use an $5\times N_s$ optical lattice that one can visualize as a ladder with $N_s$ rungs. 
There is only one atom per rung and the five sites on each rung represent the 5 possible values for $m$, with $m=0$ at the center. Tunneling in the direction orthogonal to the rung is not allowed. Tunneling along the rung generates the $X$-term.
The $U$-term is created by a parabolic potential. The $Y$-term is mediated by the $1/r^6$ interactions of the Rydberg-dressed atoms \cite{zeiher2016}. These interactions were chosen to be {\it attractive} and favoring ferromagnetism: neighbor atoms 
with the same $m$ are closer to each other than atoms with different $m$'s. By taking the distance between the rungs $a_s$ larger than the distance between the sites on the rungs $a_r$, it is possible to  do perturbation in $a_r/a_s$ in Pythagoras theorem and approximately generate the quadratic $Y$-term. A more complete discussion and illustrations 
can be found in Ref. \cite{prl121}. 

\subsection{ CARA simulators} 

In the following we discuss the possibility of adapting the idea of Ref. \cite{prl121}  to configurable arrays of Rydberg atoms 
\cite{51qubits,endres2016,keesling2019,cong2021hardwareefficient,semeghini2021probing} denoted CARA. 
They can be configured by positioning $^{87}Rb$ atoms separated by controllable (but not too small) distances, homogeneously coupled to the excited Rydberg state $\ket{r}$ with a detuning $\Delta$. The ground state is denoted $\ket{g}$ 
and the two possible states  $\ket{g}$ and $\ket{r}$ can be seen as a qubit. 
The Hamiltonian reads 
\beq
\label{eq:genryd}
H = \frac{\Omega}{2}\sum_i(\ket{g_i}\bra{r_i} + \ket{r_i}\bra{g_i})-\Delta\sum_i  n_i +\sum_{i<j}V_{ij}n_in_j,
\enq
with 
\beq\label{eq:oneosix}
V_{ij}=\Omega R_b^6/r_{ij}^6,\enq
for a distance $r_{ij}$ between the atoms labelled as $i$ and $j$. 
This repulsive interaction prevents two atoms close enough to each other to be in the $\ket{r}$ state. This is the
so-called blockade mechanism. 

This setup has been successfully used to simulate the Kibble-Zurek mechanism for chiral clock models \cite{keesling2019}. 
It has been used to propose simulators for other gauge theories 
\cite{PhysRevX.10.021041,Celi:2019lqy,PhysRevResearch.2.013288}.
Simulations with Rydberg atoms are reviewed in Ref. \cite{Wu:2020axb}. 

\begin{figure}
  \includegraphics[width=8.6cm,angle=0]{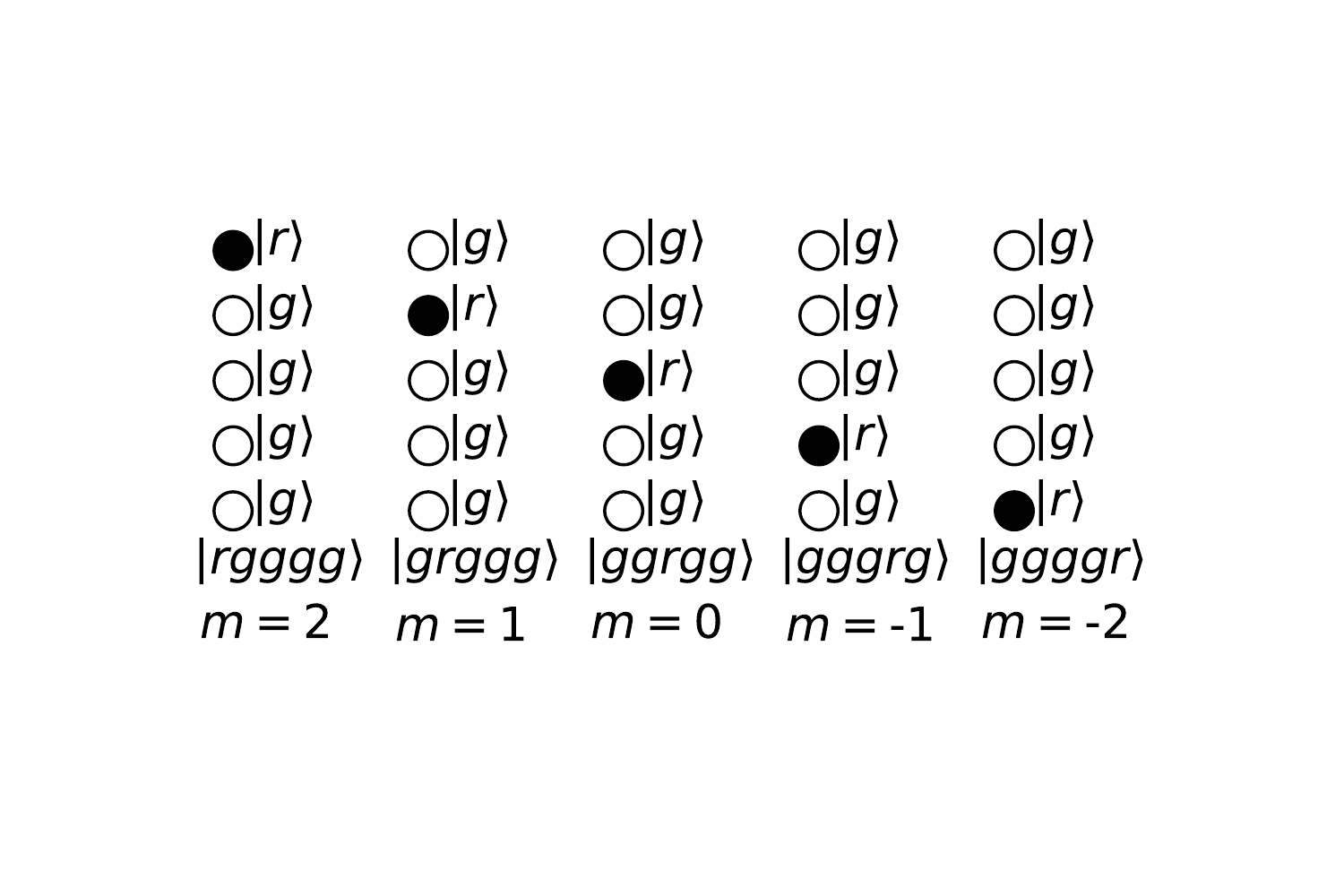}  
  \caption{ \label{fig:5lad}The 5 spin-2 states with a 5 atom setup.}
\end{figure}

In subsection \ref{subsec:ladder}, we discussed a setup \cite{prl121} where one direction of the optical lattice was used to carry the $2S+1$ spin degrees of freedom on the sites of the rungs. We will now try to replace each rung by a line of $2S+1$ Rydberg atoms close enough to each other to prevent more than one atom to be in the $\ket{r}$ state. The spin-2 case is illustrated in Fig. \ref{fig:5lad}. For instance, $m=2$ corresponds to $\ket{rgggg}$. Also notice that charge conjugation is implemented by a reflection about the horizontal axis passing by the $m=0$ states. 

With current technology, it is difficult to get 5 atoms on a line close enough to make the blockade mechanism efficient.
However it seems possible to do it with three atoms as in a spin-1 truncation. In the following we will concentrate on this simpler realization. 
 \subsection{Spin-1 CARA implementations}
 \label{subsec:2ra}
\begin{figure}[h]
  \includegraphics[width=8.6cm,angle=0]{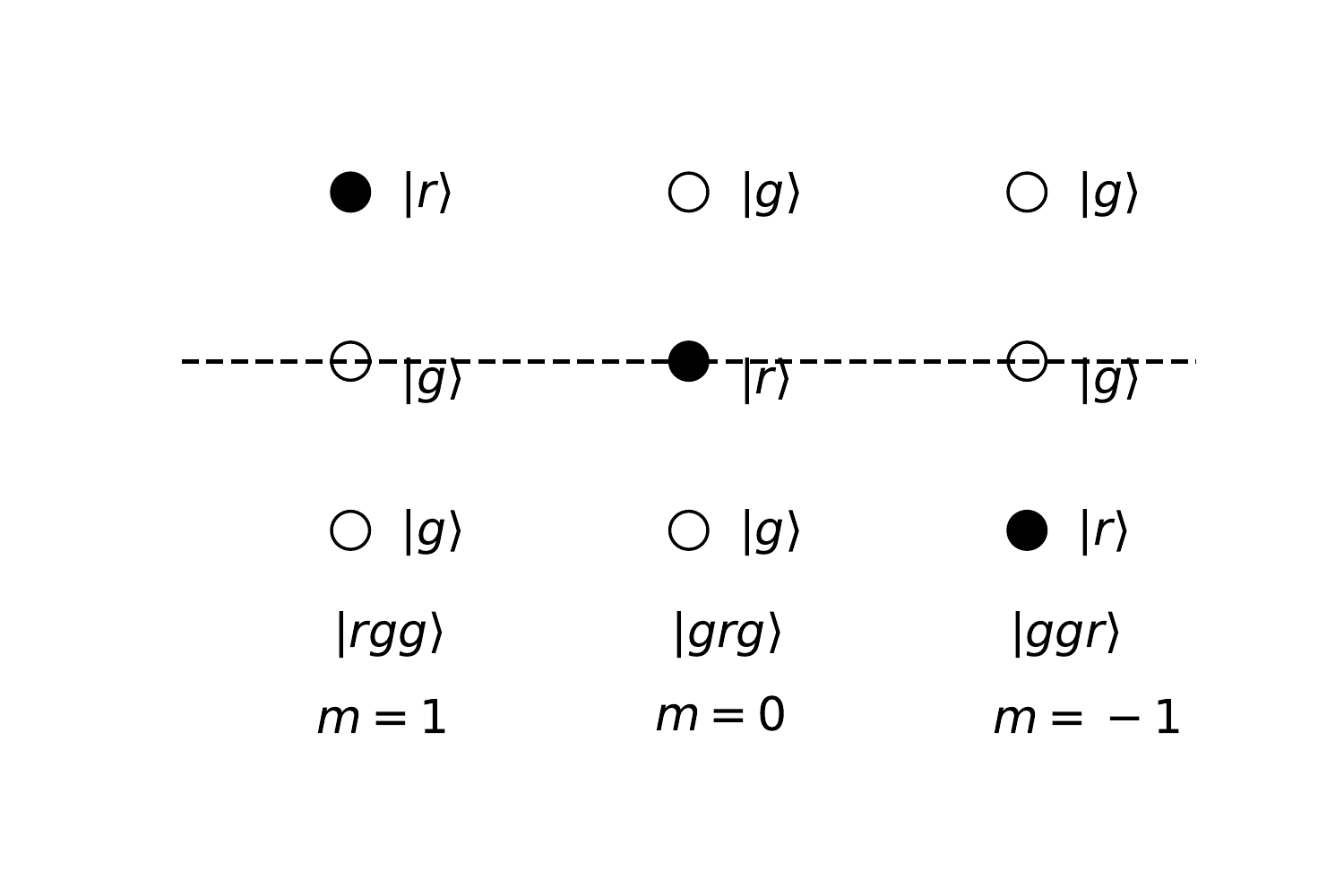}  
  \caption{ \label{fig:3lad}Three spin-1 states with the three atom setup, the 5 other possible states are not displayed.}
\end{figure}

For spin-1, we propose the correspondence 
\begin{align}
&\ket{1} \rar \ket{rgg},\nonumber\\
&\ket{0} \rar \ket{grg},\\
&\ket{-1} \rar \ket{ggr}.\nonumber
\end{align}
This is illustrated in Fig. \ref{fig:3lad}. Again, charge conjugation is implemented as a reflection with respect to the horizontal axis. 

\begin{figure}[h]
  \includegraphics[width=8.6cm,angle=0]{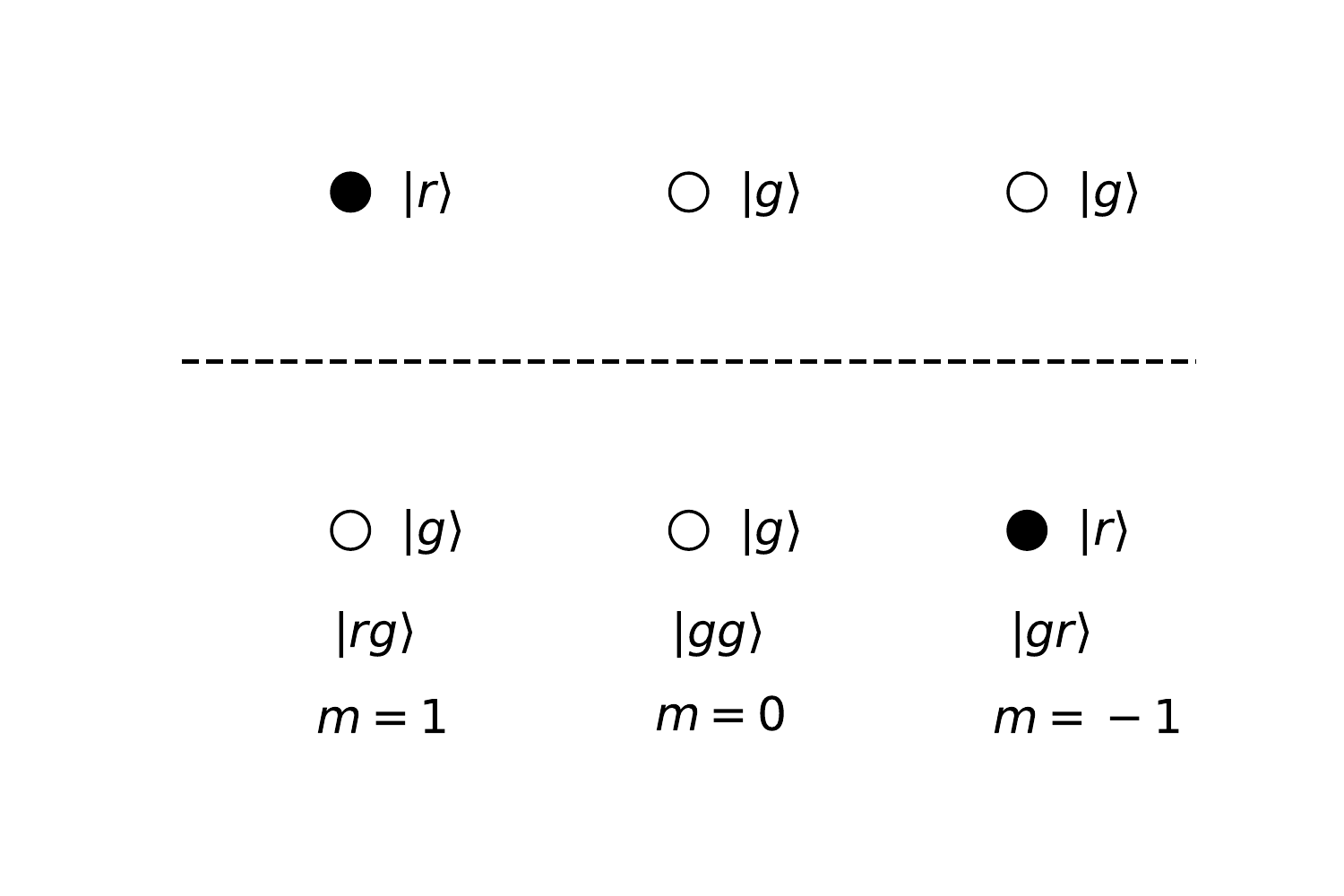}  
  \caption{ \label{fig:2lad}Spin-1 states with the two atom setup.}
\end{figure}

An even simpler setup consists in using only two Rydberg atoms and having $\ket{m=0} $ to be a state without Rydberg states, namely
\begin{align}
&\ket{1} \rar \ket{rg},\nonumber\\
&\ket{0} \rar \ket{gg},\\
&\ket{-1} \rar \ket{gr}.\nonumber
\end{align}
This is illustrated in Fig. \ref{fig:2lad}. 
Since this second possibility is the simplest, it will be the starting point of the presentation of the next sections.

\section{One spin system}
\label{sec:onespin}
In this section, we discuss the one spin-1 Hilbert space and the local Hamiltonian for the target model and the CARA implementations with two and three atoms.
\subsection{Target model}
The local part of the target Hamiltonian is 
\beq
H^{1T}=\frac{U}{2}(L^z)^2-XU^x.
\enq
We will discuss the spectrum of this Hamiltonian with emphasis on the symmetries in order to build matching simulators. 
Since $H^{1T}$ is invariant under charge conjugation, we introduce the $\mc$ eigenstates
\beq
\ket{\pm}\equiv \frac{1}{\sqrt{2}}(\ket{1}\pm\ket{-1}).
\enq
with $\mc$-eigenvalues $\pm1$: 
\beq\mc \ket{\pm}=\pm\ket{\pm}.\enq
They are also eigenstates of $(L^z)^2$ with eigenvalue 1. 
Note that 
\beq
L^z\ket{\pm}=\ket{\mp}.
\enq
In addition $\mc \ket{0}=\ket{0}$.

There is only one $\mc$-odd state which is $\ket{-}$. It is annihilated by $U^x$
\beq
U^x\ket{-}=0.
\enq
Consequently,
\beq
H^{1T}\ket{-}=\frac{U}{2}\ket{-},
\enq
for any value of $X$. 

In the $\mc$-even sector, we have 
\begin{align}
\label{eq:uaction}
&U^x\ket{0}=\frac{1}{\sqrt{2}}\ket{+},\\
&U^x\ket{+}=\frac{1}{\sqrt{2}}\ket{0},
\end{align}
and the eigenvalues are obtained from the even matrix in the $\ket{0},\ \ket{+}$ basis: 
\beq
H^{1T}_{even}=
\begin{pmatrix}
0&-\frac{X}{\sqrt{2}}\\
-\frac{X}{\sqrt{2}}&\frac{U}{2}
\end{pmatrix}.
\enq
The two eigenstates are
\beq
\ket{0}_X=\cos\phi\ket{0}+\sin\phi \ket{+},
\enq
with 
eigenvalue 
\beq
E_0(X)=\frac{1}{4}(U-\sqrt{U^2+8X^2} \ ),
\label{eq:32}
\enq
and 
\beq
\ket{+}_X=\cos\phi\ket{+}-\sin\phi \ket{0},
\enq
with 
eigenvalue 
\beq
E_+(X)=\frac{1}{4}(U+\sqrt{U^2+8X^2} \ ).
\enq
The mixing angle obeys the equation
\beq\label{eq:mixing} \tan\phi=-\sqrt{2\ }\frac{E_0(X)}{X}.\enq

If we treat $X$ as a perturbation, we obtain that at the lowest nontrivial order 
\begin{align}
E_0(X)&\simeq -\frac{X^2}{U},\\
E_+(X)&\simeq \frac{U}{2}+\frac{X^2}{U},\\
\label{eq:38}
\phi&\simeq \sqrt{2}\frac{X}{U}.
\end{align}
These results can also be derived  using standard perturbative formulas \cite{sakurai}. The perturbation has no diagonal element and the energy corrections occur at second order.

\subsection{Two Rydberg atom implementation}

The two atom setup discussed in Sec. \ref{subsec:2ra} provides a very simple implementation of a single spin-1 system. 
Introducing $\pm 1$ labels for the top and bottom atoms, we call $n_{\pm1}$ the occupation of their $\ket{r}$ state. 
The list of the possible states and their occupations are given in Table \ref{tab:2ra}.
\def\np{n_{+1}}
\def\nm{n_{-1}}
\def\uo{\frac{U}{2}}
\def\no{n_0}
\def\do{\Delta _0}
\def\vo{V_0}
\def\vop{\frac{V_0}{64}}
\begin{table}[h]
\begin{tabular}{ |c|c|c|c|c|c| } 
\hline
Setup&Ket&$n_{+1}$&$n_{-1}$&Short &Energy ($\Omega=0)$\\
 \hline
 $\begin{matrix}
\bullet\\
\circ
\end{matrix}$&$\ket{rg}$&1&0& $\ket{1}$ & $-\Delta$ \\ 
\hline
 $\begin{matrix}
 \circ\\
\circ
\end{matrix}$&$\ket{gg}$&0&0& $\ket{0}$ & 0 \\ 
\hline
 $\begin{matrix}
\circ\\
\bullet
\end{matrix}$&$\ket{gr}$&0&1& $\ket{-1}$ & $-\Delta$ \\ 
\hline
 $\begin{matrix}
\bullet\\
\bullet
\end{matrix}$&$\ket{rr}$&1&1& $\ket{2}$ & $-2\Delta+\vo$ \\ 
 \hline
\end{tabular}
\caption{\label{tab:2ra}Graphical representation of the two atoms in space, the symbol $\circ$ represents the ground state $\ket{g}$ and $\bullet$ the Rydberg state $\ket{r}$, ket notation, occupations, short notation and energy for $\Omega=0.$}
\end{table}
The Hamiltonian for the physical two-atom system is 
\begin{align}
\label{eq:2ra}
H^{2R} = &-\Delta(\np+\nm)+\vo\np\nm \\&+\frac{\Omega}{2}\sum_{\pm 1}(\ket{g_{\pm 1}}\bra{r_{\pm 1}} + \ket{r_{\pm 1}}\bra{g_{\pm 1}}).
\end{align}
We want to match this Hamiltonian with the target $H^{1T}$. 
We first consider the first term of the target Hamiltonian. The splitting between $\ket{0}$ and $\ket{\pm 1}$ is $\uo$ when $\Omega=0$, and can 
be implemented by setting 
\beq
\Delta=-\uo.
\enq
In addition we want to suppress transitions to the $\ket{rr}$ state by introducing a large enough energy $V_0$. 
This is the blockade mechanism. This can be achieved by 
positioning the two atoms close enough to each other. 

\begin{figure}
  \includegraphics[width=8.5cm]{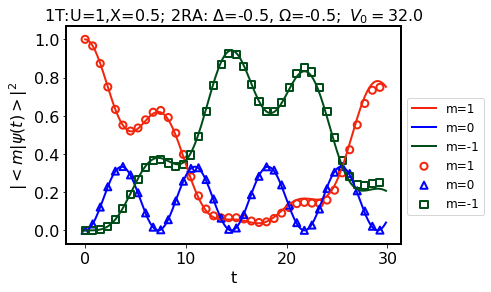}  
  \includegraphics[width=8.5cm]{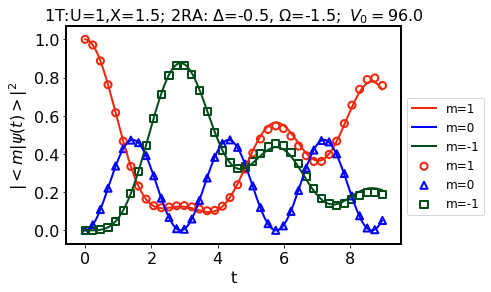}  
  \caption{\label{fig:ex1}$|\bra{m}U(t)\ket{m=1}|^2$, one site with exact Hamiltonian $U=1,\ X=0.5$ (solid lines) and Rydberg Hamiltonian with $\Omega=-0.5$, $\Delta=-0.5$  and $V_0=64|\Omega |=32$ (empty symbols) (top), and  
  $U=1,\ X=1.5$ (solid lines) and Rydberg Hamiltonian $H^{2R}$ with $\Omega=-1.5$, $\Delta=-0.5$  and $V_0=64|\Omega |=96$ (empty symbols) (bottom). }
\end{figure}

In order to determine $\Omega$, we compare the action of 
\beq
H_1^{2R}=\frac{1}{2}\sum_{\pm 1}(\ket{g_{\pm 1}}\bra{r_{\pm 1}} + \ket{r_{\pm 1}}\bra{g_{\pm 1}})
\enq
on the simulator Hilbert space to the action of $U^x$ on the target Hilbert space. 
\begin{align}
H_1^{2R}\ket{gg}&=\frac{1}{2}(\ket{rg}+\ket{gr}),\\
H_1^{2R}\frac{1}{\sqrt{2}}(\ket{rg}+\ket{gr})&=\frac{1}{\sqrt{2}}(\ket{gg}+\ket{rr}),\\
H_1^{2R}\frac{1}{\sqrt{2}}(\ket{rg}-\ket{gr})&=0.
\end{align}
Comparing with the action of $U^x$ on the three states of the target Hilbert space, we see that, except for the heavy state $\ket{rr}$, they are identical. 
Consequently, we can set 
\beq
\Omega=-X.
\enq
Except for possible transitions to $\ket{rr}$, the correspondence is exact and the linear formula applies for arbitrary values of $X$. The good matching for $U=1$ with $X=0.5$ and  $U=1$ with $X=1.5$ is demonstrated in Fig. \ref{fig:ex1}.
\subsection{Three Rydberg atom simulator}
In the three atom setup, we have an additional atom which is associated with the $m=0$ state. We label its $\ket{r}$ occupation $\no$. The Hamiltonian reads
\begin{align}
H^{3R} =&-\do \no -\Delta\sum_{j=0,\pm 1} n_j \\&+ \vo(\no\np +\no\nm) 
+V_0'\np\nm\\& +\frac{\Omega}{2}\sum_{j=0,\pm 1}(\ket{g_j}\bra{r_j} +\ket{r_j}\bra{g_j}),
\end{align}
with 
\beq
V'_0= \frac{V_0}{64}
\enq
when the three atoms are located equidistantly  on a line as in Fig .  \ref{fig:3lad}.
The spin-1 sector is shown in Table \ref{tab:spin13ra}.
\def\ci{$\circ$}
\def\bu{$\bullet$}
\begin{table}
\begin{tabular}{ |c|c|c| } 
\hline
Setup&Ket &Energy ($\Omega =0$)\\
 \hline
 $\begin{matrix}
\bullet\\
\circ\\
\circ
\end{matrix}$& $\ket{1}$ & $-\Delta$ \\ 
\hline
 $\begin{matrix}
 \circ\\
\bullet\\
\circ
\end{matrix}$& $\ket{0}$ & $-\Delta-\do$ \\ 
\hline
 $\begin{matrix}
 \circ\\
\circ\\
\bullet
\end{matrix}$& $\ket{-1}$ & $-\Delta$ \\ 
 \hline
\end{tabular}
\caption{\label{tab:spin13ra} Spin-1 states for the three atom simulator.}
\end{table}
The auxiliary sector has five states shown in Table \ref{tab:aux3ra}.
\begin{table}
\begin{tabular}{ |c|c|c| } 
\hline
Setup&Ket &Energy ($\Omega =0$)\\
 \hline
 $\begin{matrix}
\circ\\
\circ\\
\circ
\end{matrix}$& $\ket{0''}$ & 0 \\ 
\hline
 $\begin{matrix}
 \bullet\\
\bullet\\
\circ
\end{matrix}$& $\ket{1'}$ & $-2\Delta-\do+\vo$ \\ 
\hline
 $\begin{matrix}
 \bullet\\
\circ\\
\bullet
\end{matrix}$& $\ket{0'}$ & $-2\Delta+\vop$ \\ 
\hline
 $\begin{matrix}
 \circ\\
\bullet\\
\bullet
\end{matrix}$& $\ket{-1'}$ & $-2\Delta-\do+\vo$ \\ 
\hline
 $\begin{matrix}
 \bullet\\
\bullet\\
\bullet
\end{matrix}$& $\ket{3}$ & $-3\Delta-\do+2\vo+\vop$ \\ 
 \hline
\end{tabular}
\caption{\label{tab:aux3ra} Auxiliary states for the three atom simulator. }
\end{table}
Following previous notation, we also define 
\beq
\ket{\pm '}\equiv \frac{1}{\sqrt{2}}(\ket{1'}\pm\ket{-1'}), 
\enq
\def\h13{H^{3R}_1}
\def\ono{\frac{1}{\sqrt{2}}}
and 
\beq
\h13=\frac{1}{2}\sum_{j=0,\pm 1}(\ket{g_j}\bra{r_j} +\ket{r_j}\bra{g_j}),
\enq
Unlike the previous situation with two atoms, $\h13$ only connects the spin-1 sector with the auxiliary sector 
\begin{align}
\h13\ket{0}&=\frac{1}{2}\ket{0''}+\ono \ket{+'},\\
\h13\ket{+}&=\frac{1}{2}\ket{+'}+\ono \ket{0'}+\ono \ket{0''},\\
\h13\ket{-}&=\frac{1}{2}\ket{-'}.
\end{align}
Using the corresponding matrix elements together with standard perturbation theory \cite{sakurai}, 
we obtain the perturbative matching equation for the energy differences:
\begin{align}
\frac{X^2}{U}&=\frac{\Omega ^2}{2}(\frac{1}{\Delta-\vop}-\frac{1}{\Delta}),\\
\uo+\frac{X^2}{U}&=\do+\frac{\Omega ^2}{4}(\frac{1}{\Delta+\do}+\frac{2}{\vo-\Delta}-\frac{1}{\vo-\Delta-\do}).\nonumber
\end{align}
Similarly we can try to match perturbative expressions for the mixing angle $\phi$ defined in Eq. (\ref{eq:mixing}). However, $\phi$ 
contributions appear at first order in $X$ in the target model (because $U^x$ connects $\ket{0}$ and $\ket{+}$), but only at second 
order in $\Omega$ in the three-atom simulator. More explicitly, 
\beq
\sqrt{2}\frac{X}{U}=\frac{\Omega^2}{2\sqrt{2}\do}(\frac{1}{\Delta}+\frac{1}{\vo-\Delta-\do}).
\enq
This apparently contradicts the idea that $X$ and $\Omega$ should be proportional for small values of $X$. 
If we are given $U$ and $X$, we have three nonlinear equations for $\Omega$, $\Delta$ , $\do$ and $\vo$ and generically, 
we expect one-parameter families of solutions. By fixing one of the unknowns, one can look for solutions using Newton's method. More generally, it is easy to find accurate numerical solutions for the energies and mixing of the simulators and 
it seems possible to attack the matching problem non perturbatively. In order to pursue such effort, it would be useful to know the range of values corresponding to feasible experiments. 

With today's technology, it seems difficult to create non homogeneous detuning and we should consider the limit where $\do$ is zero. 
In this limit, $\ket{0}$ and $\ket{+}$ are degenerate when $\Omega$ is set to zero. In the $\ket{0}$ and $\ket{+}$  basis, 
the energy matrix up to second order in $\Omega$ has the form 
\beq
-\Delta\mathbb{1}-\frac{\Omega^2}{4} \mathbb{M},
\enq
with 
\beq
\mathbb{M}=
\begin{pmatrix}
\frac{1}{\Delta}+\frac{2}{\vo-\Delta}&\frac{\sqrt{2}}{\Delta}+\frac{\sqrt{2}}{\vo-\Delta} \\
\frac{\sqrt{2}}{\Delta}+\frac{\sqrt{2}}{\vo-\Delta}&\frac{2}{\Delta}+\frac{1}{\vo-\Delta}-\frac{2}{\Delta-\vop}
\end{pmatrix}.
\enq
The matrix $\mathbb{M}$ determines the mixing angle and the eigenvalues. 

\subsection{An example of approximate solution with three Rydberg atoms}
\begin{figure}
  \includegraphics[width=8.5cm]{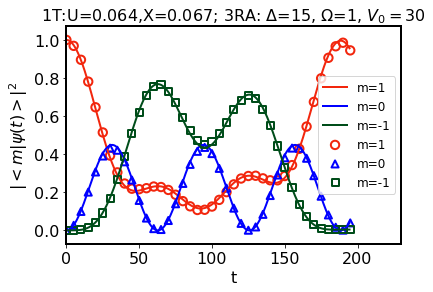}  
  \caption{\label{fig:ex2}$|\bra{m}U(t)\ket{m=1}|^2$ for one site with exact Hamiltonian $U=0.064,\ X=0.067$ (solid lines), Rydberg Hamiltonian with $\Omega=1$, $\Delta=15$  and $V_0=30$ (empty symbols).  }
\end{figure}

The three-atom simulator leads to nonlinear equations. However, it is not difficult to find 
approximate solutions when $\Delta$ and $V_0$ are in a specific ratio. 
As a simple example, we picked
\beq \do=0 \ {\rm and}\ \vo=2\Delta. 
\enq 
If we neglect the $\vop$ term, we have 
\beq
\mathbb{M}\simeq \frac{1}{\Delta}\begin{pmatrix} 3&2\sqrt{2}\\2\sqrt{2}\ &1\end{pmatrix}.
\enq
Given that there is an overall minus sign in front of $\mathbb{M}$, its largest eigenvalue corresponds to the lowest energy state and the mixing angle approximately satisfies the equation 
\beq
\tan \phi\simeq\frac{1}{\sqrt{2}}
\enq
Comparing with Eq. (\ref{eq:mixing}), we find that this angle corresponds to the situation $X=U$. 
Note that for this significant value $X/U=1$, the angle is not small and the linear approximation of Eq. (\ref{eq:mixing}) given in Eq. (\ref{eq:38}) is not accurate.
Instead, we used the exact value of $E_0$ given in Eq. (\ref{eq:32}). For the simulator,  
this large mixing angle is  {\it not} controlled by
$\Omega^2$. This is a feature of degenerate perturbation theory. This procedure is justified by the good quality of the agreement 
between target and simulator shown in Fig. \ref{fig:ex2}.

Furthermore, we can compute the energy spectrum in this simple example. The eigenvalues of $\mathbb{M}$ are 
approximately $5/\Delta$ and $-1/\Delta$. In addition, we have up to second order in $\Omega$:
\beq
E_-=-\Delta -\frac{\Omega^2}{4}\frac{1}{\vo-\Delta}.
\enq
Consequently, in this simple example, we have 
\begin{align}
E_+-E_0&=\frac{3}{2}\frac{\Omega^2}{\Delta},\\
E_--E_0&=\frac{\Omega^2}{\Delta}.
\end{align}
In the target model with $X=U$, we have 
\begin{align}
E_+-E_0&=\frac{3}{2}U,\\
E_--E_0&=U.
\end{align}
The ratio of differences are both 3/2 and we can match the scale for $\Omega^2/\Delta=U$. 
Given that the matrix has been approximated, we should look for matching in the region $X\simeq U$.  
As a numerical example, we used  $\Omega=1$, $\Delta=15$  and $V_0=30$ and found good matching for $U=0.064,\ X=0.067$ both close to $1/15=0.0666...$
This is illustrated in Fig. \ref{fig:ex2}. Note that the time scale in $1/\Omega$ units is significantly larger than in the two atom case. This is due to the extra $\Omega/\Delta$ factor in the energy scale. 

\section{Two-spin system}
\label{sec:twospins}
In this section, we follow the same sequence as in Sec. \ref{sec:onespin} for a two-spin system
motivated by the Hamiltonian of Eq. (\ref{eq:ham}).
\subsection{Target model}
For the target model we consider two spins called left (L) and right (R) connected by a $Y$-term.   
The target Hamiltonian for the two-spin system is chosen to be
\begin{align}
H^{2T}=&H^{1T}_L+H^{1T}_R+\frac{Y}{2}(L_L^z-L_R^z)^2\\
=&\frac{U}{2}((L_L^z)^2+(L_R^z)^2)-X(U_R^x +U_L^x )+\frac{Y}{2}(L_L^z-L_R^z)^2\nonumber
\end{align}
 For the time evolution we consider an initial state $\ket{0,0}$, and calculate the probability to stay in that state or to be in the state 
\beq
\label{eq:sym}
\ket{S}\equiv\frac{1}{2}(\ket{0,1}+\ket{0,-1}+\ket{1,0}+\ket{-1,0}).
\enq
 obtained by applying the $X$-term on the initial state. This guarantees a significant overlap when $X$ is not too small. 
 For the comparison of the four and six atom simulators we picked a special target situation where solutions of the one-spin problems are available, more specifically we picked 
$U=1$, $X=1.2$ and $Y=0.2$.
\subsection{Four Rydberg atom simulator}
\begin{figure}
  \includegraphics[width=8.6cm,angle=0]{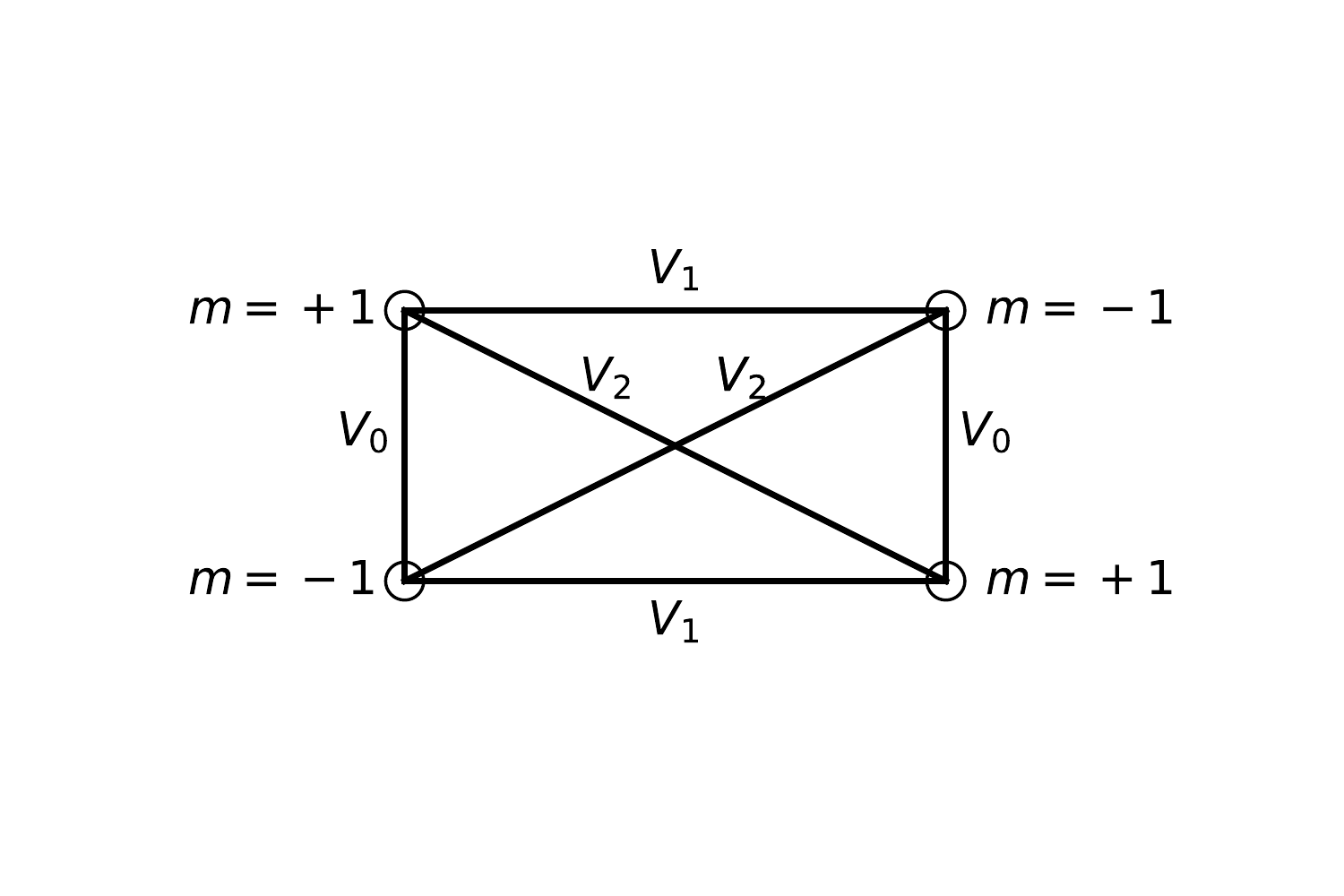}  
  \caption{ \label{fig:4rav} Rydberg interactions for the four atom simulator.}
  \end{figure}

For a simulator with four atoms, we use the two-atom setup for two pairs and include the additional $V_{ij}$ appearing in Eq. (\ref{eq:genryd}). The Hamiltonian reads
\begin{align}
H^{4R} =&H^{2R}_L+H^{2R}_R\nonumber \\&+ V_1(n_{+1L} n_{-1R}+n_{-1L} n_{+1R})\\
&+V_2(n_{+1L} n_{+1R}+n_{-1L} n_{-1R})\nonumber
\end{align}
When the $V_i $ are positive, we need to have the opposite signs closer (bipartite charge conjugations on alternate sites).
In other words, if the interactions are repulsive and decreasing like $1/r^6$, we need to put atoms with the same $m$ farther apart. This is illustrated in Fig. \ref{fig:4rav}. 
For the standard $1/r^6$ Rydberg interactions as in Eq. (\ref{eq:oneosix}), and using $a_r$ and $a_s$ as the vertical (as in one spin) and horizontal (coupling the two spins)  lattice spacings respectively, 
and their ratio
\beq
\rho\equiv \frac{a_r}{a_s},
\enq
we have
\begin{align}
\label{eq:v12}
V_1=&V_0\rho^6,\\
V_2=&V_0(\rho/\sqrt{1+\rho^2}\ )^6.
\end{align}
The matching condition when $X=\Omega=0$ reads
\begin{align}
\Delta&=-\uo-\frac{Y}{2},\\
V_1&=Y,\\
V_2&=-Y.
\end{align}
The second equation can be solved using 
\beq
\rho=(\frac{Y}{V_0})^{1/6}.
\enq
The third equation has no solutions for  positive $V_2$. In Eq. (\ref{eq:v12}) $V_2$ is smaller than $V_1$ but nevertheles positive. This implies that it is not possible to exactly match the energy of the states $\ket{+1,-1}$ and $\ket{-1,+1}$. 
The possible solutions to this problem are: 1) ask experimentalists to adjust the couplings locally, 2) consider $Y$-perturbations 
over situations with non-zero $\Omega$. On the other hand, the phase structure and dynamical features of the simulator are worth exploring even if the 
matching with the target is not perfect. 
Note that for the six-atom setup  discussed below, we will see that the matching in the limit $\Omega=X=0$ is approximately possible following the mechanism invoked in Ref. \cite{prl121}. 

\begin{figure}[h]
  \includegraphics[width=8.5cm]{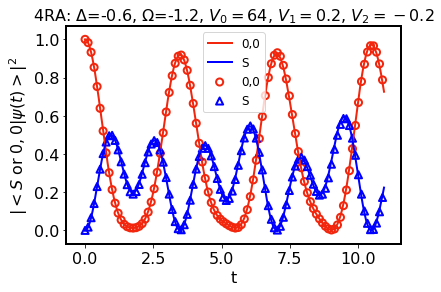}  
 \includegraphics[width=8.5cm]{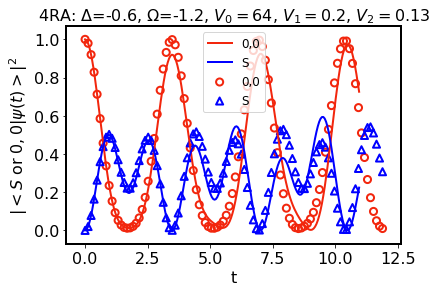}  
 \caption{\label{fig:caram}$|\bra{0,0}U(t)\ket{0,0}|^2$ and $|\bra{S}U(t)\ket{0,0}|^2$, with $\ket{S}$ defined in Eq. (\ref{eq:sym}),  for  the target two-spin Hamiltonian with $U=1$, $X=1.2$ and $Y=0.2$ (solid lines) and the four atom simulator with 
 $\Delta$=-0.6, $\Omega$=-1.2, $V_0=64$, $V_1=0.2$, with $V_2=-0.2$ (top) and $V_2=0.13$ (bottom). }
\end{figure}

In order to give an idea of the size of the effects discussed above for the four atom system, we have considered the problem mentioned in the target section, first with 
the futuristic $V_2=-Y$ (the agreement is excellent) and then with $V_2$ is as in Eq. (\ref{eq:v12}) (deviations from exact are quite visible for $t\gtrsim 3$). The results are displayed in Fig. \ref{fig:caram}. 
\subsection{Six  Rydberg atom implementation}
\begin{figure}
  \includegraphics[width=8.6cm,angle=0]{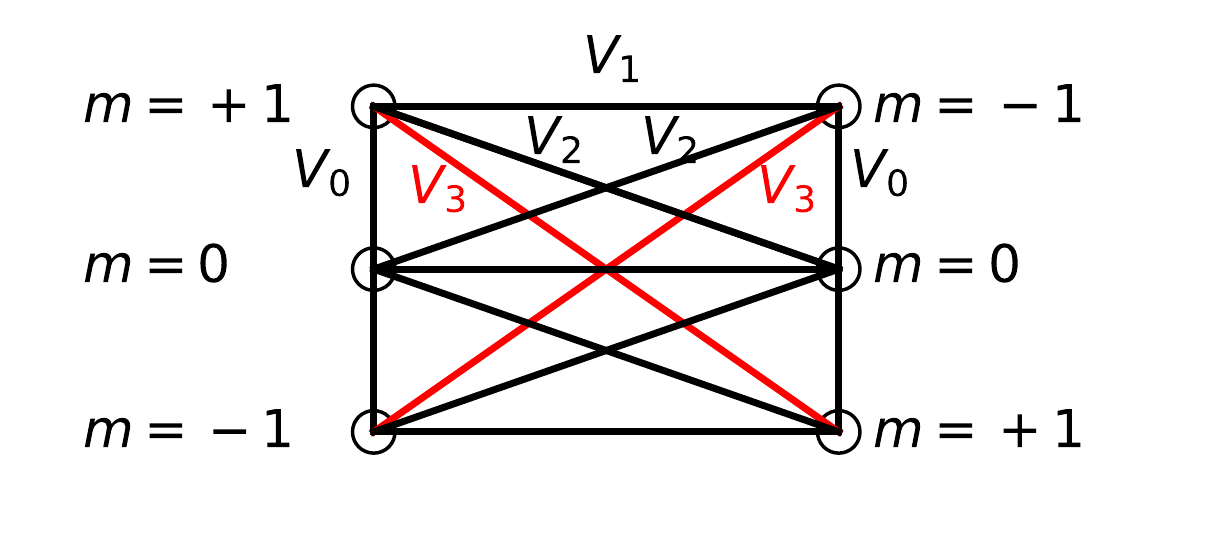}  
  \caption{ \label{fig:6rav}Rydberg interactions for the six atom simulator.}
  \end{figure}

\begin{figure}
  \includegraphics[width=8.5cm]{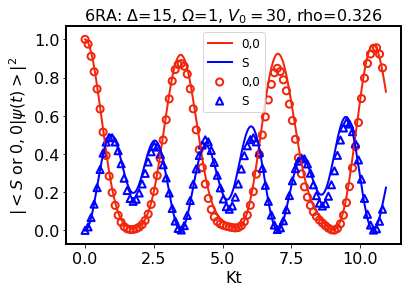} 
  \caption{\label{fig:caram6}$|\bra{0,0}U(t)\ket{0,0}|^2$ and $|\bra{S}U(t)\ket{0,0}|^2$, with $\ket{S}$ defined in Eq. (\ref{eq:sym}),  with the target two-spin Hamiltonian with $U=1$, $X=1.2$ and $Y=0.2$ (solid lines) and the six atom simulator with 
$\Delta$=15, $\Omega$=1, $V_0=30$, $\rho$=0.326. A rescaling $K= 0.05464$ has been applied to the simulator time to match the target evolution.}
\end{figure}

We now consider a six-atom setup with two three-atom spin-1 setups.  The Hamiltonian reads
\begin{align}
H^{6R} =&H^{3R}_L+H^{3R}_R\nonumber \\&+ V_1(n_{+1L} n_{-1R}+n_{-1L} n_{+1R})\nonumber\\
&+V_2(n_{0L}(n_{+1R} +n_{-1R}))\\
&+V_2((n_{+1L} +n_{-1L})n_{0R})\nonumber \\
&+V_3(n_{+1L} n_{+1R}+n_{-1L} n_{-1R})\nonumber
\end{align}
When the $V_i $ are positive, we again need to have the atoms representing spins of opposite signs closer in space. This is illustrated in Fig. \ref{fig:6rav}. 
  For the standard Rydberg interactions of Eq. (\ref{eq:oneosix}), we have $V_1$ and $V_2$ as in Eq. (\ref{eq:v12}) and 
\beq
V_3=V_0(\rho/\sqrt{1+4\rho^2}\ )^6
\enq
Matching condition when $X=\Omega=0$ are
\begin{align}
\Delta_0&=\uo+Y\nonumber \\
V_1-V_2&=\frac{Y}{2}\\
V_1-V_3&=2Y,\nonumber
\end{align}
These equations have approximate solutions with a tuned $\rho$ 
when $Y$ is not too large. In Fig. \ref{fig:caram6}, we used $\Delta$=15, $\Omega$=1, $V_0=30$, and empirically tuned $\rho$ to a value 0.326. A rescaling $K= 0.05464$ has been applied to the simulator time to match the target evolution. In Fig. \ref{fig:caram6}, we show the case 
$\Delta$=15, $\Omega$=1, $V_0=30$, as in the one spin-1, with  $\rho$=0.326. 
 
 
\section{Digital implementations} 
\label{sec:ibm}
So far we have discussed an analog simulator with a computational basis given by the states $\ket{g}$ and $\ket{r}$ for each atom. However, it is also possible to calculate the time evolution corresponding to the Rydberg atom Hamiltonians $H^{nR}$ 
using digital methods, for instance a universal quantum computer with qubits. Before going farther, it should be mentioned that a large $V_0$ requires small Trotter steps (of order $1/V_0$) in order to suppress the contributions with multiple $\ket{r}$ states by fast oscillations. This is not optimal with NISQ machines and it would be profitable to find a native way to implement the blockade. However, we will discuss the results to give an idea of the time scales involved. 

\begin{figure}
  \includegraphics[width=8.5cm]{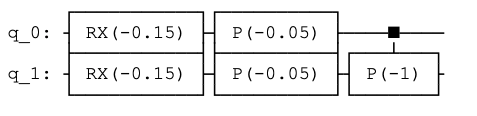}  
 \caption{\label{fig:circuit}Qiskit circuit for a Trotter form of the evolution operator for $H^{2R}$ described in the text .}
 \end{figure}

\begin{figure}
  \includegraphics[width=8.5cm]{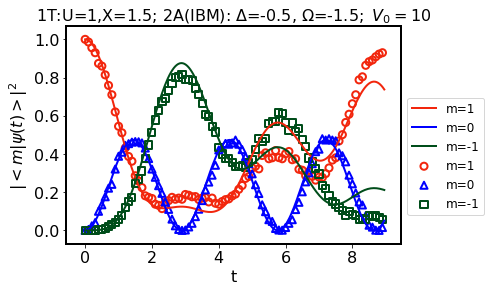}  
  \caption{\label{fig:exIBM}$|\bra{m}U(t)\ket{m=1}|^2$, one site with exact Hamiltonian   $U=1,\ X=1.5$ (solid lines) and Rydberg Hamiltonian $H^{2R}$ with $\Omega=-1.5$, $\Delta=-0.5$  and $V_0=10$, with the circuit of Fig. \ref{fig:circuit} calculated with the Qiskit simulator (empty symbols). }
\end{figure}

As a simple example, we considered the two atom Hamiltonian $H^{2R}$ with the values $\Omega=-1.5$, $\Delta=-0.5$ as in Fig. \ref{fig:ex1},  but with a lower value $V_0=10$. We used a Trotter step $\delta t=0.1=1/V_0$. We used Qiskit with the  classical simulator called with the instruction 
\verb|Aer.get_backend('qasm_simulator')| and which sample the so-called statevector \cite{sqiskit}. This simulator does not include a realistic noise model corresponding to current hardware such as, for instance,  \verb|ibmq_lima|.
The circuit is shown in Fig. \ref{fig:circuit} with Qiskit notations. We used 1,000 shots for each of the times. 
 In Fig. \ref{fig:circuit}, $P$ is the phase gate 
  \beq
  P(\phi)\equiv 
  \begin{pmatrix}
  1&0\\
  0&\exp(i\phi)
  \end{pmatrix},
  \enq
 and we have the usual definition
  \beq
 RX(\lambda)\equiv \exp(-i\frac{\lambda}{2 }X). \enq
The results of the Qiskit simulations are shown in Fig. \ref{fig:exIBM}.
The Trotter errors become quite significant when $t\gtrsim 2$. Based on recent runs on IBMQ \cite{bench}, we would expect  that computations on actual IBMQ hardware 
with a slightly larger Trotter step would lead to reasonable results for $t\lesssim 1$. 
The two spin system can be implemented with universal quantum computers provided that an all-to-all connectivity is available. 
Hardware considerations will be discussed in a separate paper. 

\section{Conclusions}
\label{sec:conclusions}

In summary, we have proposed a ladder-shaped CARA with two and three atoms for a single spin-1 and four and six atoms 
for two coupled spins. In one spatial dimension this is all we need to control in order to study larger systems. 
We compared target and simulators using perturbative and numerical methods. The two-atom setup provides an easily controllable simulator of the one-spin model while the three-atom setup involves a nonlinear matching. 
However when two spins are coupled, the six-atom simulator provides solutions closer to the target than the four-atom simulator for small $\Omega$. 
Approximate implementations of the two-spin model appear to be  possible with near term technology. 
Extensions to spin-2 and implementations in 2+1 dimensions are under investigation.

\newpage
{\it Acknowledgments}. This work is supported in part by the U.S. Department of Energy (DoE) under Award Number DE-SC0019139. 
Special thanks to Alex Keesling who enthusiastically supported the idea using the Rabi driven term to replace the tunneling in the orginal formulation and made very interesting suggestions. We also thank  Shan-Wen Tsai, Jin Zhang, Johannes Zeiher, James Corona, Erik Gustafson, Denis Candido, Michael Flatte, Craig Pryor, Nate Gemelke, Tout Wang, Shangtao Wang and 
the members of QuLAT for suggestions and comments.
%

\end{document}